\documentclass[a4paper]{article}

\usepackage[pages=all, color=black, position={current page.south}, placement=bottom, scale=1, opacity=1, vshift=5mm]{background}
\SetBgContents{
	\tt This work is shared under a \href{https://creativecommons.org/licenses/by-nc/4.0/deed.en}{CC BY-NC 4.0 license}.
}      % copyright

\usepackage[margin=1in]{geometry} 

\usepackage{amsmath}
\usepackage{amsthm}
\usepackage{amssymb}
\usepackage{booktabs}

\usepackage{float}

\usepackage[ruled, lined, longend]{algorithm2e}

\usepackage[utf8]{inputenc}
\usepackage{hyperref}
\hypersetup{
	unicode,
	pdfauthor={Author One, Author Two, Author Three},
	pdftitle={A simple article template},
	pdfsubject={A simple article template},
	pdfkeywords={article, template, simple},
	pdfproducer={LaTeX},
	pdfcreator={pdflatex}
}

\usepackage{multirow}
\usepackage{caption}
\usepackage{subcaption}
\usepackage{amsmath}
\usepackage{amsfonts}
\usepackage{graphicx}

\theoremstyle{plain}

\theoremstyle{definition}

\title{Solving Heterogeneous Agent Models with \\ Physics-informed Neural Networks}
\author{Marta Grześkiewicz \\ University of Cambridge}
\date{Work in Progress \\ This version: November 2025}

\begin{document}

\maketitle

\begin{abstract}
Understanding household behaviour is essential for modelling macroeconomic dynamics and designing effective policy. While heterogeneous agent models offer a more realistic alternative to representative agent frameworks, their implementation poses significant computational challenges, particularly in continuous time. The Aiyagari-Bewley-Huggett (ABH) framework, recast as a system of partial differential equations, typically relies on grid-based solvers that suffer from the curse of dimensionality, high computational cost, and numerical inaccuracies. This paper introduces the ABH-PINN solver, an approach based on Physics-Informed Neural Networks (PINNs), which embeds the Hamilton-Jacobi-Bellman and Kolmogorov Forward equations directly into the neural network training objective. By replacing grid-based approximation with mesh-free, differentiable function learning, the ABH-PINN solver benefits from the advantages of PINNs of improved scalability, smoother solutions, and computational efficiency. Preliminary results show that the PINN-based approach is able to obtain economically valid results matching the established finite-difference solvers. We hope this will open new avenues for solving complex heterogeneous agent models in macroeconomics.
\end{abstract}

\section{Introduction}

Understanding household behaviour is a central challenge in economics. As the fundamental decision-making units in an economy, households collectively drive macroeconomic dynamics, shaping everything from business cycle fluctuations to long-run growth. Accurate models of consumption and saving are essential for policy design, welfare analysis, and forecasting. Historically, these models relied heavily on the representative agent paradigm, in which a single, aggregate agent stands in for the entire population. While computationally convenient, this approach fails to capture distributional dynamics and underperforms in predicting behaviour in the tails of income or wealth distributions.

The emergence of heterogeneous agent models (HAMs) marked a major theoretical advance. By allowing for persistent idiosyncratic shocks and incomplete markets, these models better reflect the micro-level frictions and risk exposures that shape aggregate outcomes. The canonical Aiyagari-Bewley-Huggett (ABH) framework formalises these ideas, embedding household heterogeneity in a general equilibrium setting. Individuals face uninsurable income shocks and solve intertemporal optimisation problems under borrowing constraints, giving rise to endogenous, stationary distributions of wealth and consumption. This framework has since become foundational for modern macroeconomics, particularly the class of Heterogeneous Agent New Keynesian (HANK) models, which integrate micro-level frictions into general equilibrium models of monetary and fiscal policy.

Recent advances by \cite{achdou2022income} have further reformulated the ABH model in continuous time, recasting the household problem as a coupled system of non-linear partial differential equations (PDEs): a Hamilton-Jacobi-Bellman (HJB) equation characterising individual optimisation, and a Kolmogorov Forward (KF) equation describing the evolution of the distribution of agents across states. This is solved by a grid-based finite-difference (FD) solver. While this reformulation offers greater analytical elegance and compatibility with modern policy tools, it also introduces substantial numerical challenges. These PDE systems are difficult to solve analytically and rely on computationally intensive grid-based methods for approximation.

Grid-based solvers, though widely used, face three fundamental limitations. First, they suffer from the curse of dimensionality: the computational cost scales exponentially with the number of state variables. Even modest extensions to the ABH model, such as adding housing, human capital, or debt, can render the problem intractable \cite{ahn2018inequality}. Second, these methods impose a high computational burden, especially in dynamic policy environments where models must be repeatedly solved under varying parameters \cite{fernandez2023financial, moll2014productivity}. Third, they are prone to discretisation error and interpolation inaccuracies, and require complex, error-prone implementations that can obscure economic insights \cite{achdou2022income, judd1998numerical}.

To address these limitations, we propose a novel solution method based on Physics-Informed Neural Networks (PINNs) \cite{RAISSI2019686}, the ABH-PINN solver. PINNs are a class of universal function approximators that embed the governing PDEs of a model directly into the training objective of a neural network. Originally developed in computational physics, they have demonstrated success in solving high-dimensional, non-linear systems where traditional methods fail. By leveraging the structure of the economic model itself, (in this case, the HJB and KF equations) PINNs transform the solution process from brute-force grid search to an elegant function approximation task that is both mesh-free and differentiable by construction.

In the context of HAMs, PINNs offer several advantages. They alleviate the curse of dimensionality by learning over randomly sampled collocation points rather than fixed grids, making the inclusion of additional state variables computationally feasible. They significantly reduce computational costs by enabling rapid evaluation of policy functions once trained, which is especially valuable in the repeated-solution contexts typical of policy simulations. They also provide smooth, continuous approximations to value and distribution functions, avoiding the interpolation and discretisation issues of grid-based methods. Crucially, they are well-suited to handle non-linearities, occasionally binding constraints (e.g., borrowing limits), and sparse or noisy data, both common features of macroeconomic models.

In this paper, we present preliminary results to demonstrate the applicability of PINNs to solving the ABH model in continuous time with the ABH-PINN solver, with a view to benchmarking it against the FD solver presented in \cite{achdou2022income} and to show how it performs with higher dimensions of heterogeneity in a future iteration of this work. The remainder of the paper is structured as follows. Relevant literature is presented in section 2, and section 3 presents the theoretical background. Section 4 presents the ABH-PINN solver, with preliminary results presented in section 5. We conclude and discuss future work in section 6.

\section{Relevant Literature}

\subsection{PINNs in Economics and Finance}

To the best of our knowledge, PINNS have so far been applied to problems in macroeconomics only in \cite{hu2024new}, which applies a PINN to solve the spatial Solow growth model. In this setting, the PINN is trained to predict the evolution of capital and technology over space and time. The model is benchmarked against traditional numerical approaches, including the spatial Solow framework with non-concave production functions proposed by \cite{capasso2010spatial}. PINNs yield strong performance, achieving error rates of 3.4\% (maximum) and 0.5\% (minimum) in capital and technology levels. Additionally, visual comparisons with results from \cite{kabanikhin2020differential}, who apply differential evolution techniques to similar models, suggest that PINNs can successfully replicate both homogeneous and heterogeneous equilibrium distributions, including bimodal patterns. Although quantitative error metrics are not provided for this comparison, qualitative agreement supports the validity of the PINN approach.

The inverse problem of inferring model parameters from data is more challenging but of critical importance for empirical economics and policy analysis. \cite{hu2024new} demonstrates that PINNs can accurately recover parameters with minimal data: errors remain below 3\% with only two known data points, and fall under 1\% with 30 data points, even under moderate noise. This robustness to data sparsity and noise is particularly relevant in macroeconomics and econometrics, where observational data is often incomplete or imprecise. Crucially, simulated settings with known ground truth allow for rigorous validation of the method’s inference accuracy.

PINNs have been applied more extensively in computational finance. Early work by \cite{tanios2021physics}, in a master’s thesis, initiated the use of PINNs for option pricing under the Black-Scholes and Heston models, particularly in high-dimensional asset spaces. This foundational study opened the door to further innovation in neural network architectures tailored for financial applications.

Building on this foundation, \cite{bai2022application} introduced a suite of architectural enhancements to improve accuracy and training stability. These include local adaptive activation functions and a slope recovery term in the loss function—techniques that enhance convergence and precision. Their framework was applied not only to standard Black-Scholes pricing but also to more complex PDEs, such as the Ivancevic model, which includes rogue wave and soliton solutions.

Recent studies further highlight the breadth of PINN applications in finance. PINNs are applied to option pricing in \cite{hainaut2024option}, who validate the effectiveness of PINNs in pricing options under the Heston model, which incorporates stochastic volatility, and in \cite{gatta2023meshless}, who explore PINNs as a meshless alternative for pricing American options, which involve a free boundary due to the early exercise feature. Traditional grid-based methods struggle in this context, whereas PINNs offer a more flexible and potentially more accurate solution. Finally, \cite{nuugulu2024physics} extend the PINN framework to solve time-fractional PDEs under the Time-Fractional Black-Scholes Equations incorporating memory effects and long-range dependence, which is a phenomena often observed in financial time series but poorly captured by classical models.

\subsection{Heterogeneous Agent Models in Macroeconomics}

The study of macroeconomic phenomena has undergone a profound transformation with the rise of HAMs, marking a significant departure from the long-dominant representative agent framework. While those models simplify economic analysis by assuming all agents are identical, HAMs explicitly account for the diverse circumstances of individual households and firms. This shift provides a richer, more empirically grounded lens for analysing economic aggregates, inequality, and the heterogeneous effects of policy interventions.

The intellectual groundwork for modern HAMs was laid by seminal papers that introduced idiosyncratic risk, incomplete markets, and endogenous wealth distributions into general equilibrium analysis. These models provided the first coherent framework for understanding how rational, forward-looking individuals behave when they cannot fully insure against uncertainty. \cite{BEWLEY1977252} initiated this line of inquiry by analysing a single consumer facing uninsurable, random income fluctuations and an explicit borrowing constraint. His work formally established the motive for precautionary saving, demonstrating that individuals accumulate assets to self-insure against potential future hardship. This insight challenged the permanent income hypothesis by showing how liquidity constraints and uncertainty systematically alter consumption-saving decisions.

This framework was extended to a general equilibrium setting in \cite{HUGGETT1993953}. In an endowment economy populated by infinitely-lived, \textit{ex-ante} identical agents, he demonstrated how idiosyncratic income risk leads to an endogenous distribution of wealth. The collective desire for precautionary savings by all agents serves to increase the aggregate supply of capital, thereby depressing the equilibrium risk-free interest rate below what would prevail in a complete markets setting. Huggett's model provided a powerful mechanism for explaining both the observed level of wealth inequality and the equity premium puzzle. \cite{aiyagari1994uninsured} subsequently embedded these insights into a standard neoclassical growth model with production. In his framework, the aggregate capital stock is the sum of assets held by a continuum of heterogeneous households facing uninsurable labour income shocks. Aiyagari demonstrated that the presence of incomplete markets could lead to a higher aggregate saving rate and a larger capital stock compared to the representative-agent benchmark. The model generates a stationary distribution of wealth, providing a crucial link between micro-level behaviour and macroeconomic aggregates. Although these foundational models successfully introduced key mechanisms, they systematically underestimated the extreme concentration of wealth observed in empirical data \cite{de2015quantitative}.

Building on this foundational work, subsequent research introduced greater realism to better match empirical facts, particularly the high degree of wealth inequality. \cite{krusell1998income} made a major stride by incorporating aggregate risk (i.e., business cycle fluctuations) into the Aiyagari framework. A key challenge in such an environment is that the entire distribution of wealth becomes a state variable, making the model computationally intractable. Their critical contribution was to show that the aggregate dynamics could be accurately approximated using only a few moments of the wealth distribution, most notably the mean. This discovery of ``approximate aggregation'' made it feasible to study business cycles within heterogeneous agent environments. \cite{de2004wealth} addressed the failure of standard models to generate realistic wealth concentration at the top of the distribution. By introducing intergenerational transfers through voluntary and accidental bequests, she showed that savings motives related to passing on wealth are crucial for explaining the emergence of large fortunes and matching the thick upper tail of the wealth distribution observed in the data. More recently, \cite{kaplan2014model} refined the understanding of household consumption behaviour by developing a two-asset model. They distinguish between a liquid asset (e.g. cash) and an illiquid, higher-return asset (e.g., housing or retirement accounts). This framework generates a class of ``wealthy hand-to-mouth'' households—agents who hold substantial illiquid wealth but very few liquid assets. These households exhibit a high Marginal Propensity to Consume (MPC) out of transitory income shocks, a feature with profound implications for the transmission of economic policy.

The insights from the HAM literature have been integrated with standard New Keynesian models, which feature nominal rigidities, to create HANK models. These models have become the dominant framework for modern monetary and fiscal policy analysis. HANK models, pioneered in works like \cite{kaplan2018monetary}, demonstrate that heterogeneity is of first-order importance for the transmission of macroeconomic policy. In contrast to Representative Agent New Keynesian models where the transmission of monetary policy works primarily through inter-temporal substitution, HANK models highlight powerful indirect effects that operate through general equilibrium changes in labour income. Because many households have high MPCs, changes in interest rates that affect firm investment and labour demand have large and immediate impacts on consumption. This channel significantly amplifies the power of monetary policy. Similarly, these models provide a clear framework for analysing fiscal stimulus, as they predict that transfers to low-liquidity, high-MPC households will have the largest impact on aggregate demand \cite{ravn2017job}.

A recent and significant development in the field has been the shift from discrete-time to continuous-time modelling. Although early HAMs were almost exclusively formulated in discrete time, this approach has computational and theoretical disadvantages. As argued by \cite{achdou2022income}, a continuous-time frameworks allow for richer dynamics, more precise characterisation of equilibrium, and the use of powerful mathematical tools. The modern continuous-time approach formulates the household's problem as a HJB partial differential equation and the evolution of the wealth distribution as a KF equation. The coupling of these two equations fully characterises the aggregate equilibrium. This mathematical structure draws inspiration from Mean Field Game theory \cite{lasry2007mean}, which studies strategic decision-making in large populations of interacting agents. The seminal work of \cite{achdou2022income} provided a comprehensive theoretical characterisation and a fast, reliable numerical solution method (based on finite differences) for the canonical ABH model in continuous time. This breakthrough has made continuous-time methods widely accessible, spurring the development of a new generation of continuous-time HANK models for sophisticated policy analysis (e.g., \cite{auclert2021using}). These new methods allow for a more granular and computationally efficient analysis of complex economic dynamics, pushing the frontier of macroeconomic research.

\section{Background}

\subsection{ABH Economy} \label{section:abh}

In this section, we present the ABH model, presented in \cite{achdou2014heterogeneous} which is a more general formulation of \cite{achdou2022income}, by including their extension to a general income process, and assuming the closing of the economy as in \cite{aiyagari1994uninsured} that wealth takes the form of productive capital that is used by a representative firm which also hires labour. The micro-foundations consist of individual optimisation decisions, based on their current labour and capital income and expenditure on consumption. 

\subsubsection{Setup}

\paragraph{Individuals.} There is a continuum of individuals that are heterogeneous in their wealth and labour productivity, earning them an income based on their wages and labour supply. The state of the economy is the joint distribution of wealth and productivity. Each individual agent has standard preferences and optimises their utility 
    \begin{equation} \label{eq:u-max}
        \mathbb{E}_0 \int_0^\infty e^{-\rho t} u(c_t) \text{d}t
    \end{equation}
where $u(c_t)$ represents utility from consumption at time $t$, $c_t$, with $u$ as a strictly increasing and strictly concave function. $\rho \geq 0$ is the discount rate and $\mathbb{E}_0$ is the expectation operator at the initial time period $t = 0$. Their wealth, $a_t$, takes the form of bonds and evolves according to the differential equation, given by
\begin{equation} \label{eq:wealth_de}
    \text{d} a_t = [w_t z_t + r_t a_t - c_t] \text{d}t
\end{equation}
where $w_t$ is their wage and $z_t$ their productivity, making $w_t z_t$ is labour income, and $r_t$ is the interest rate, making $r_t a_t$ the capital income. The individuals also face a borrowing limit 
\begin{equation} \label{eq:borrowing-constraint}
    a_t \geq \underline{a}, \quad \text{where } -\infty < \underline{a} \leq 0.
\end{equation}
Labour productivity evolves over time as a stationary diffusion process
\begin{equation} \label{eq:labour_diffusion}
    \text{d} z_t = \mu_z(z_t) \text{d}t + \sigma_z(z_t) \text{d}W_t.
\end{equation}
The individuals maximise \ref{eq:u-max} subject to \ref{eq:wealth_de}, \ref{eq:borrowing-constraint}, and \ref{eq:labour_diffusion}, taking as given the interest rate and wage.

\paragraph{Firms.} There is a representative firm with a production function $Y=F(K,L)$ exhibiting constant returns to scale. The total amount of capital is determined by the total amount of wealth (bonds), given by
\begin{equation} \label{eq:agg-K}
    K(t) = \int \int a g(a, z, t) \, \text{d}a \, \text{d}z
\end{equation} 
and we normalise the total amount of labour supplied in the economy to one. With competitive markets and $\delta$ as the rate of depreciation of capital, the interest rate and wage are given by 
\begin{align}
    r(t) &= \partial_K F(K(t), 1) - \delta \\
    w(t) &= \partial_L F(K(t), 1)
\end{align}
with the shorthand notation $\partial_K F = \partial F / \partial K$ and $\partial_L F = \partial F/\partial L$. 

\subsubsection{Recursive formulation}

The evolution of the economy through the consumption-savings decisions of individuals and the joint distribution of their wealth and labour productivity is summarised by two PDEs: a Hamilton-Jacobi-Bellman (HJB) equation and a Kolmogorov Forward (KF) (or Fokker-Planck) equation. Given an initial distribution of $g$ at time $t=1$ and appropriate boundary conditions, the two PDEs with the equilibrium relationship described in the firm equations fully characterise the evolution of the economy. 

\paragraph{HJB equation.} The HJB equation is given by
\begin{equation} \label{eq:hjb}
    \begin{split}
    \rho v(a, z, t) = \max_c \bigg\{ & u(c) + \partial_a v(a, z, t)[w(t)z + r(t)a - c] + \partial_z v(a, z, t)\mu_z(z) \\
        & + \frac{1}{2} \partial_{zz} v(a, z, t) \sigma_z^2(z) + \partial_t v(a, z, t) \bigg\}
\end{split}
\end{equation}
where $\rho v(a, z, t)$ represents the discounted lifetime utility an individual can obtain; $v(a, z, t)$ denotes the value function, which obtains the value the individual gets in terms of utility over the `state space', in this model defined by wealth, productivity and time. Utility from instantaneous consumption is given by $u(c)$; $\partial_av(a, z, t)$ represents the sensitivity of the value function with respect to wealth while $[w(t)z + r(t)a - c]$ represents the change in wealth (labour and capital income minus consumption). Together, $\partial_av(a, z, t)[w(t)z + r(t)a - c]$ intuitively represent the marginal value of saving, combining the expected change in value arising from wealth drift, weighted by the marginal value of wealth. It can also be interpreted as the utility gained from saving an additional unit of wealth a. $\mu_z(z)$ is the productivity drift term, and it weighs the sensitivity of the value function to labour productivity $\partial_zv(a,z,t)$. Intuitively higher productivity influences personal income and therefore the value function, but its impact will vary based on the level of wealth and time. $\frac{1}{2}\partial_{zz}v(a,z,t)\sigma_z^2(z)$ reflects \ref{eq:labour_diffusion}. Future productivity is uncertain, leading to income volatility that cannot be insured against. The higher the uncertainty, the greater the weight placed on additional wealth which functions as self-insurance. $\partial_tv(a,z,t)$ represents the change in the value function with respect to time, incorporating changes to wage trends, interest rate shifts and horizon effects in finite models.

\textit{Boundary condition.} The borrowing constraint is not present in the HJB and it only binds on the boundary
\begin{equation}
    \partial_a v(\underline{a},z,t) \geq u'(\mu_a(\underline{a}, z,t))
\end{equation}
which binds at $a = \underline{a}$. See \cite{achdou2022income} for the derivation. 

\paragraph{KF equation.} The KF equation is given by
\begin{align}
\partial_t g(a, z, t) = & -\partial_a [\mu_a(a, z, t) g(a, z, t)] - \partial_z [\mu_z(z) g(a, z, t)] \\
& + \frac{1}{2} \partial_{zz} [\sigma_z^2(z) g(a, z, t)]
\end{align}
 
It utilises the savings policy function $\mu_a(a, z, t) = w(t)z + r(t)a - c(a, z, t)$ from \ref{eq:hjb} to analyse the forwards and backwards movements of individuals within the feature space. $\partial_t g(a,z,t)$ gives the rate of change of the distribution over time, which is informed by $-\partial_a [\mu_a(a, z, t) g(a, z, t)]$ representing the drift of wealth. 

This drift informs the x-axis movements of the population distribution. In a simplified model with realistic specifications, you’d expect the mass of consumers to gradually shift towards the right as they accumulate wealth through saving. The second term $-\partial_z [\mu_z (z) g(a, z, t)]$ tracks drift for the idiosyncratic productivity process. Productivity diffusion is modelled through $\frac{1}{2} \partial_{zz} [\sigma_z^2(z) g(a, z, t)]$ accounting for the stochastic spread reflecting uncertainty in productivity stemming from \ref{eq:labour_diffusion}.

\subsection{PINNs}

Neural networks are often used to model data processes due to their nature as universal function approximators \cite{hornik1989multilayer}, and their ability to take advantage of automatic differentiation \cite{baydin2018automatic}. PINNs are neural networks constrained during training to respect any symmetries, invariances, or conservation principles originating from the physical laws that govern observed data, as modelled by general time-dependent and non-linear partial differential equations \cite{RAISSI2019686}. These constraints are imposed through the loss function used to train the network. 

A loss function for a PINN is a weighted sum of losses arising from the residual of the PDE, the initial conditions (IC), boundary conditions (BC) and physical constraints (phys), given by
\begin{equation}
    \mathcal{L}_{\text{total}} = \lambda_{\text{PDE}} \cdot \mathcal{L}_{\text{PDE}} + \lambda_{\text{IC}} \cdot \mathcal{L}_{\text{IC}} + \lambda_{\text{BC}} \cdot \mathcal{L}_{\text{BC}} + \lambda_{\text{phys}} \cdot \mathcal{L}_{\text{phys}}
\end{equation}
where the scalars $\lambda$ determine the contribution of the components in the loss function. The components are described as follows: 
\begin{itemize}
    \item \textbf{PDE residual loss, $\mathcal{L}_{\text{PDE}}$.} Evaluating the trained network comparing to the PDE yields a \textit{residual}: zero residual means the network exactly satisfies the governing law. 

    \item \textbf{Initial-condition loss, $\mathcal{L}_{\text{IC}}$.} Initial condition can be arrived at by either knowing the initial profile of a system and comparing it to the output of the PINN, or by solving for the steady-state solution and comparing for backward problems.

    \item \textbf{Boundary-condition loss, $\mathcal{L}_{\text{BC}}$.} At the boundaries of the domain we may require enforcement of edge restrictions, so the learned solution behaves sensibly (e.g. zero-flux). 

    \item \textbf{Physics (economic) constraints, $\mathcal{L}_{\text{phys}}$.} Beyond the governing PDE, additional shape restrictions: monotonicity of marginal utility, concavity of the value function, or conservation of total probability mass can be softly imposed. Each restriction is written as an inequality or identity that the network should satisfy. The squared violation enters $\mathcal{L}_{\text{phys}}$.
\end{itemize}

PINNs can solve two kinds of problems: firstly, \textit{given fixed model parameters $\lambda$ what can be said about the unknown hidden state $u(t,x)$ of the system?}, and secondly, \textit{what are the parameters $\lambda$ that best describe the observed data?} In this work, we focus on the former problem in the context of the ABH model, where the parameters of the PDEs are obtained through calibration, and leave the application to observed data for future work. 

\section{Methodology}

In this section, we present the PINN-based method to solve the PDEs that describe the workhorse ABH micro-founded macroeconomic model, the HJB and KF, as described in section \ref{section:abh}. We first present the general framework for the method, then describe the details for solving each step.

\subsection{General Framework}

Two PINNs approximate the solution to the ABH model. The first approximates the value function $v(a, z, t)$ governed by the HJB, and the second approximates the density function $g(a, z, t)$ governed by the KF equation. The training loss minimises the combined physics-informed residuals, initial and boundary conditions, and economic constraints. The algorithmic structure is as follows: 

\begin{algorithm}[H]
\caption{ABH-PINN Solver}
\label{alg:abh_pinn_solver}

\SetKwInOut{Input}{Input}
\SetKwInOut{Output}{Output}
\SetKwComment{Comment}{$\triangleright$\ }{}

\Comment{Initialise networks}
Initialise HJB-PINN $\hat{v}_{\theta}$ to approximate $v(a,z,t)$ \;
Initialise KF-PINN $\hat{g}_{\nu}$ to approximate $g(a,z,t)$ \;

\Comment{Pretraining}
Train $\hat{v}_{\theta}$ for a number of epochs with fixed $r$, $w$ \;

\While{not converged}{
    Update HJB-PINN $\hat{v}_{\theta}$ \;
    Compute savings policy $\mu_a$ from current HJB-PINN \;
    Update KF-PINN $\hat{g}_{\nu}$ using $\mu_a$ \;
    Compute aggregate capital $K$ from updated $\hat{g}_{\nu}(a,z,t)$ \;
    Update $r$, $w$ using new $K$ \;
}

\end{algorithm}

Pretraining is not strictly necessary but we have found it speeds up training as the KF-PINN is not updated while the value function is still in initial stages of training. The full loss is a weighted combination:
\begin{equation}
L = \lambda_{PDE}^{HJB} L_{\text{PDE}}^{\text{HJB}} + \lambda_{PDE}^{KF} L_{\text{PDE}}^{\text{KF}} + \lambda_{IC} L_{\text{IC}} + \lambda_{BC} L_{\text{BC}} + \lambda_{mass} L_{\text{mass}} + \lambda_{phys} L_{\text{phys}}.
\end{equation}

\subsection{Estimating of value function $v(a,z,t)$ from the HJB equation}

\subsubsection{PDE Loss}

The core of the HJB solution is a residual loss constructed from the PDE in equation (\ref{eq:hjb}):
\begin{equation}
    \begin{split}
        L_{\text{PDE}}^{\text{HJB}} = \mathbb{E}_{(a, z, t) \sim \mathcal{D}} \Big[ \rho \hat{v}_\theta(a, z, t) &- u(c^*(a, z, t)) - \partial_a \hat{v}_\theta(a, z, t) \mu_a^*(a, z, t) - \partial_z \hat{v}_\theta(a, z, t) \mu_z(z) \\
        &- \frac{1}{2} \partial_{zz} \hat{v}_\theta(a, z, t) \sigma_z^2(z) - \partial_t \hat{v}_\theta(a, z, t) \Big]^2
    \end{split}
\end{equation}
Optimal consumption $c^*(a,z,t)$ is computed directly using the gradients of the network since the utility function is known, which informs $\mu_a^*(\cdot)$. PINNs evaluate the residual at randomly sampled points $(a_i,z_i,t_i)$, constructing the evaluation dataset $\mathcal{D}$, offering flexibility in the number and distribution of training points. 

\subsubsection{Initial Conditions}

The initial condition loss can be used to enforce either an initial state or the terminal (steady) state of the value function. These can be known, estimated (by e.g. another PINN or another solver), or guessed to guide the network to a reasonable solution. Taking $v_{ic}$ to be the initial condition for the value function at time $t=0$, we have
\begin{equation}
L_{\text{IC}} = \mathbb{E}_{(a, z) \sim \mathcal{D}} [\hat{v}_\theta(a, z, 0) - v_{ic}(a, z)]^2.
\end{equation}

\subsubsection{Boundary Conditions}

The boundary condition loss ensures that the learned value function respects the limits of the state space. In our case, this includes a Neumann slope condition at the borrowing constraint $a \geq \underline{a}$.  As described in section \ref{section:abh}, the boundary constraint at the lower bound of wealth is given by
\begin{equation}
L_{\text{BC}}^{\underline{a}} = \mathbb{E}_{(z, t) \sim \mathcal{D}} [\partial_a \hat{v}_\theta(\underline{a}, z, t) - u'(\mu_a(\underline{a}, z, t)]^2.
\end{equation}
We also impose Neumann constraints on the upper bound of $a$, and the bounds of $z$ 
\begin{equation}
L_{\text{BC}}^{\bar{a}} = \mathbb{E}_{(z, t) \sim \mathcal{D}} [\partial_a \hat{v}_\theta(\bar{a}, z, t)]^2.
\end{equation}
\begin{equation}
L_{\text{BC}}^{\bar{z}} = \mathbb{E}_{(z, t) \sim \mathcal{D}} [\partial_z \hat{v}_\theta(a, \bar{z}, t)]^2.
\end{equation}
\begin{equation}
L_{\text{BC}}^{\underline{z}} = \mathbb{E}_{(z, t) \sim \mathcal{D}} [\partial_z \hat{v}_\theta(a, \underline{z}, t)]^2.
\end{equation}

\subsubsection{Physics (Economic) Constraints}

We enforce economic regularities for the form of the value function through an additional loss term to ensure monotonicity and concavity. Monotonicity is enforced by
\begin{equation}
L_{\text{phys}} = \mathbb{E}_{(a, z, t) \sim \mathcal{D}} \big[ \min(0, \partial_{a} \hat{v}_\theta(a, z, t)) \big]^2
\end{equation}
Under some preferences, such as CRRA, the value function is strictly concave in wealth. As such, impose restrictions on the second derivative by 
\begin{equation}
L_{\text{phys}} = \mathbb{E}_{(a, z, t) \sim \mathcal{D}} \big[ \max(0, \partial_{aa} \hat{v}_\theta(a, z, t)) \big]^2
\end{equation}
An alternative to specifying such constraints is to impose restrictions on the weights of the HJB-PINN, by, for example, an input-concave neural network as described in \cite{grzeskiewicz2025uncovering}. 

\subsection{Estimating of the density $g(z,a,t)$ from the KF equation}

Let $\phi_{\nu}(a,z,t)$ denote the output of the KF-PINN and set $\hat{g}_{\nu}(\cdot) := \text{softplus}(\phi_{\nu}(\cdot))$ to ensure that the density is strictly positive.

\subsubsection{PDE Loss}

The residual loss resulting from the KF PDE is given by
\begin{equation}
    L_{\text{PDE}}^{\text{KF}} = \mathbb{E}_{(a, z, t) \sim \mathcal{D}} \Big[ \partial_t \hat{g}_{\nu}(a, z, t) + \partial_a[\mu^*_a(a,z,t) \hat{g}_{\nu}(a, z, t)] + \partial_z[\mu_z(z) \hat{g}_{\nu}(a, z, t)]  - \frac{1}{2} \partial_{zz}[\sigma_z^2 \hat{g}_{\nu}(a, z, t)] \Big]^2
\end{equation}
where $\mu^*_a(a,z,t)$ is the savings under the optimal consumption given the value function computed by HJB-PINN. 

\subsubsection{Mass-Conservation Penalty}

To enforce the normalisation condition such that we have a proper probability density for $\hat{g}_{\nu}(a, z, t)$ the mass must integrate to one, such that 
\begin{equation}
L_{\text{mass}} = \left( \int_{a,z\sim\mathcal{D}_{mesh}} \hat{g}_{\nu}(a, z, t) \text{d}a \text{d}z - 1 \right)^2
\end{equation}
this is numerically estimated by a fine mesh grid. 

\subsection{Estimating aggregate capital $K$, and updating $r$ and $w$ }

We compute and update the aggregate $K$, as given by equation (\ref{eq:agg-K}) by generating a mesh grid, and update $r$ and $w$ based on the known production function.
    
\section{Experiments}

\subsection{Setup}

We follow \cite{achdou2014heterogeneous} for the parmeters of the model economy. 

\paragraph{Individuals.} We take the utility function to be the standard CRRA, $u(c) = \frac{c^{1-\gamma}}{1-\gamma} , \gamma>0$, with $\gamma=2$. Optimal consumption is then $c^*(a,z,t) = [v_a(a,z,t)]^{-1/\gamma}$. The process is assumed to be simple Brownian motion $\text{d}z_t = \sigma_z \text{d}W_t$ reflected at $\underline{z}$ and $\bar{z}$, where $\sigma_z=0.02$, $\underline{z}=0.5$ $\bar{z}=1.5$. and assume the discount rate to be $\rho=0.05$. We set the domain for the PDEs as $\underline{a} = 0, \bar{a}=5, \underline{z}=0.5, \bar{z}=1.5, T=10$. We assume all individuals have initial wealth distributed by a Gaussian with mean $1.0$ and standard deviation $0.2$.  

\paragraph{Firms.} The production function is assumed to have Cobb-Douglas form $F(K,L)=K^{\alpha}L^{1-\alpha}$. With labour normalised to 1, $L=1$ we have $F(K,1)=K^{\alpha}$ and the implied pricing formulas are $w = (1-\alpha)K^{\alpha}$ and $r=\alpha K^{\alpha-1}-\delta$. We assume $\alpha = 0.3, \delta=0.05$.

\subsection{Implementation}

The value function $\hat{v}_{\theta}(a,z,t)$ and the density function $\hat{g}_{\nu}(a,z,t)$ are each approximated using fully connected neural networks with $3$ hidden layers of $128$ units and hyperbolic tangent (tanh) activation functions. To ensure non-negativity of the density, the output of the KF-PINN is passed through a softplus activation. 

We assume a heuristic initial condition for the value function as $v(a,z,t=0)=\log(1+a+z^2)$, chosen for numerical convenience. This is chosen for the following reasons: it is infinitely differentiable in both $a$ and $z$, which helps automatic differentiation and improves stability when training the HJB-PINN; it is strictly increasing in  $a$ and $z$; it avoids singularities at the boundaries; and has a non-trivial gradient. No formal terminal condition is imposed at $t \rightarrow \infty$. The robustness of the result to this heuristic is left for the future iteration of this work.

We take the weights in the loss function to be $1.0$ for the initial condition losses and the mass loss, and set the rest at $0.1$. At each step, we sample $100$ points from a grid of size 11 for each of the inputs to obtain the PDE residuals. To improve stability and interpretability, the training loop is divided into two phases: in the first phase, only the HJB-PINN is trained while the interest rate and wage are fixed at initial values; in the second phase, the full equilibrium system is solved, allowing both the value and density networks to be trained jointly, and updating prices endogenously to match aggregate capital. The first phase is $2,500$ steps. We train for a total of $25,000$ steps, where for the first $7,500$ steps, we train using the Adam optimiser \cite{kingma2017adammethodstochasticoptimization}, after which we switch to SGD \cite{robbins1951stochastic}. We clip gradients to stabilise training. To reduce the computational burden, we perform the equilibrium update every $5$ steps.

\subsection{Preliminary Results}

\begin{figure}[H]
    \centering
    \includegraphics[width=\linewidth]{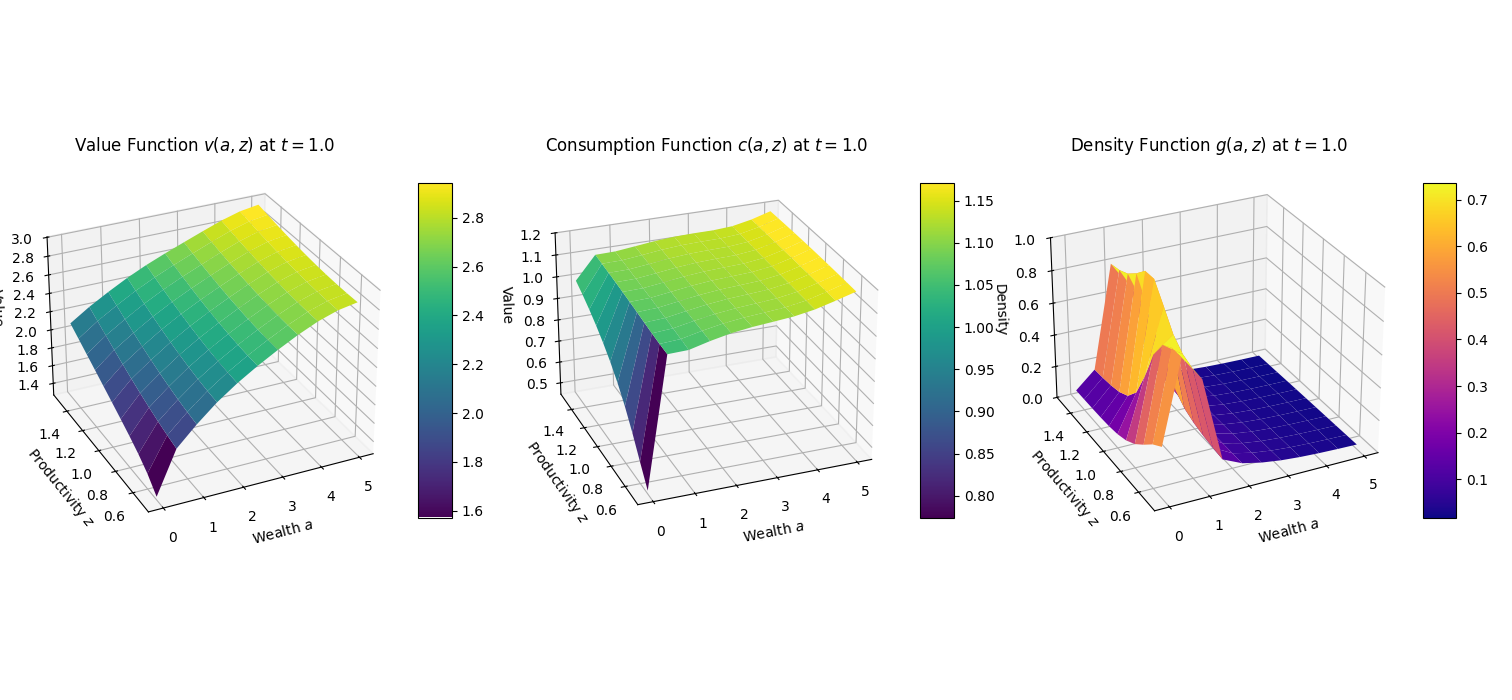}
    \caption{Value, consumption and density functions as trained by ABH-PINN at $t=1$.}
    \label{fig:t=1}
\end{figure}

\begin{figure}[H]
    \centering
    \includegraphics[width=\linewidth]{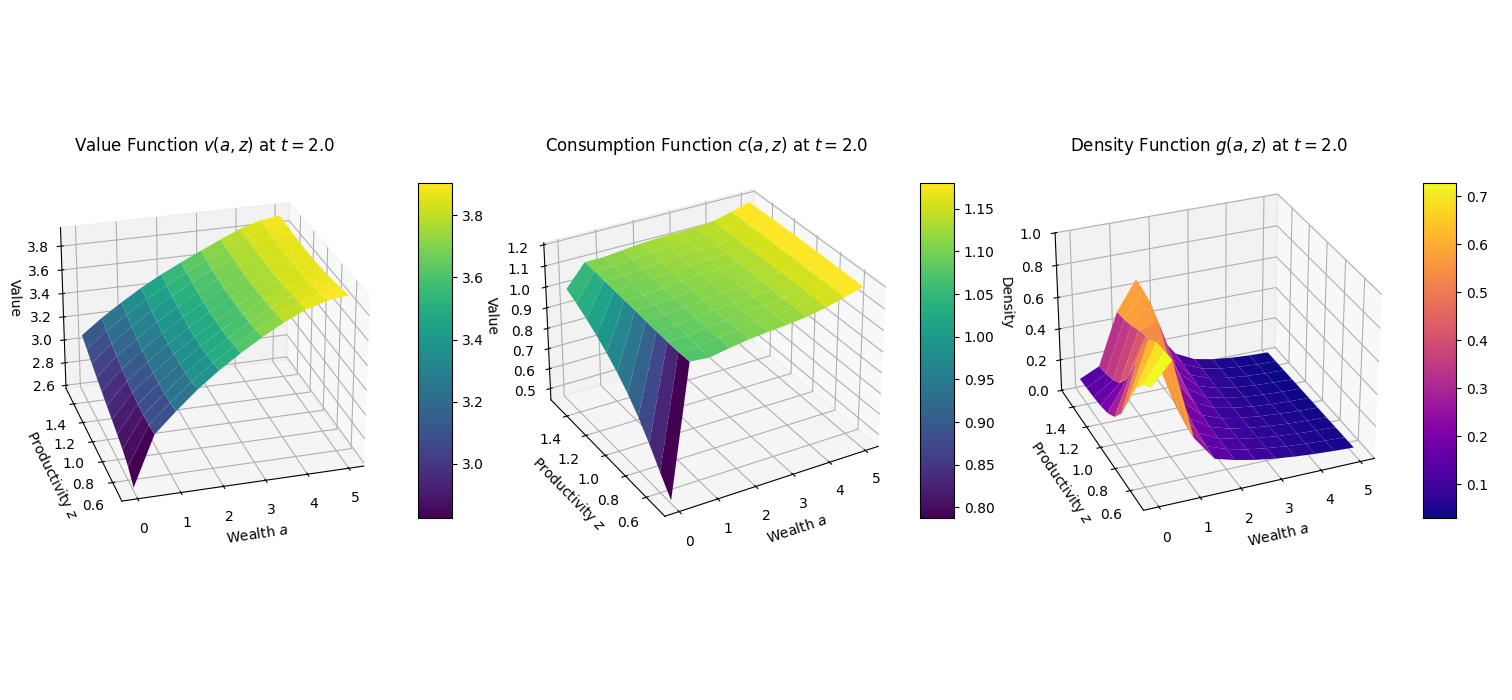}
    \caption{Value, consumption and density functions as trained by ABH-PINN at $t=2$.}
    \label{fig:t=2}
\end{figure}

\begin{figure}[H]
    \centering
    \includegraphics[width=\linewidth]{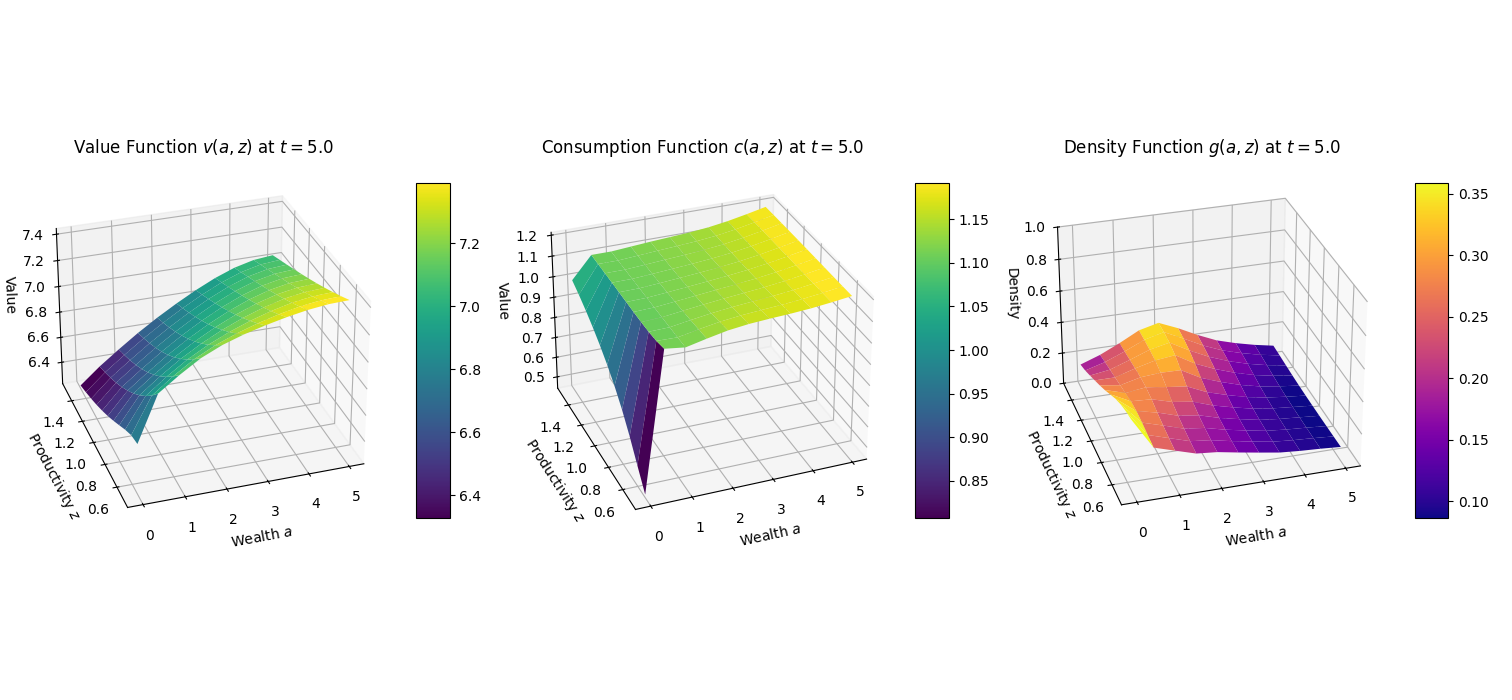}
    \caption{Value, consumption and density functions as trained by ABH-PINN at $t=5$.}
    \label{fig:t=5}
\end{figure}

\begin{figure}[H]
    \centering
    \includegraphics[width=\linewidth]{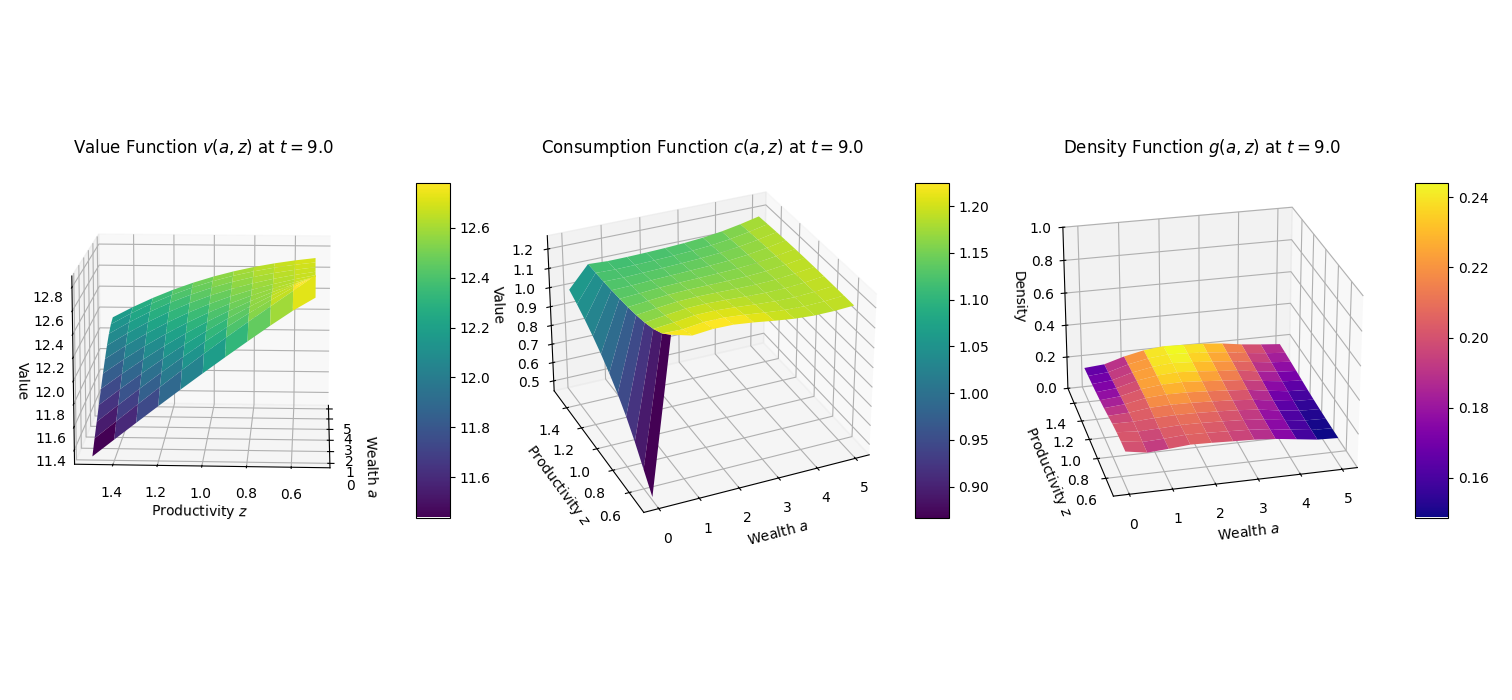}
    \caption{Value, consumption and density functions as trained by ABH-PINN at $t=9$.}
    \label{fig:t=9}
\end{figure}

We present the fitted functions for time periods 1, 2, 5, and 9 in Figures \ref{fig:t=1}, \ref{fig:t=2}, \ref{fig:t=5} and \ref{fig:t=9}, respectively. The top-left panel of each figure depicts the value function, $v(a,z, \cdot)$, which consistently shows that agent utility increases with both wealth $a$ and productivity $z$, reflecting the diminishing marginal utility of wealth through its concave shape. Correspondingly, the middle panel displays the consumption function, $c(a,z, \cdot)$, where consumption is also an increasing function of both wealth and productivity. The rightmost panel, a central element of the analysis, shows the density function, $g(a,z, \cdot)$, which captures the distribution of agents across different states. Over time, the plots for $t=1,2$, and $5$ demonstrate a clear shift in this distribution: the initial, more dispersed distribution of agents concentrates into a single, prominent peak as the economy matures. This evolution of the density function represents the aggregate outcome of individual agents optimizing their consumption and saving decisions, ultimately leading to a more concentrated and stable distribution of wealth and productivity as the economy converges toward its long-run equilibrium.

We also see that low-productivity individuals tend to be concentrated at the lower end of the wealth distribution. Since their wage income is lower due to their productivity shock, their ability to save is more limited. While the entire economy sees an overall increase in average wealth and the density function shifts to the right, a significant portion of the population, specifically those with lower productivity, remains at a lower level of accumulated wealth. The two-peak structure visible in the density plots, especially in the earlier time steps, illustrates this: one peak represents the concentration of individuals with higher productivity and wealth, while the other represents the concentration of individuals with lower productivity and wealth.

\begin{figure}[H]
    \centering
    \includegraphics[width=\linewidth]{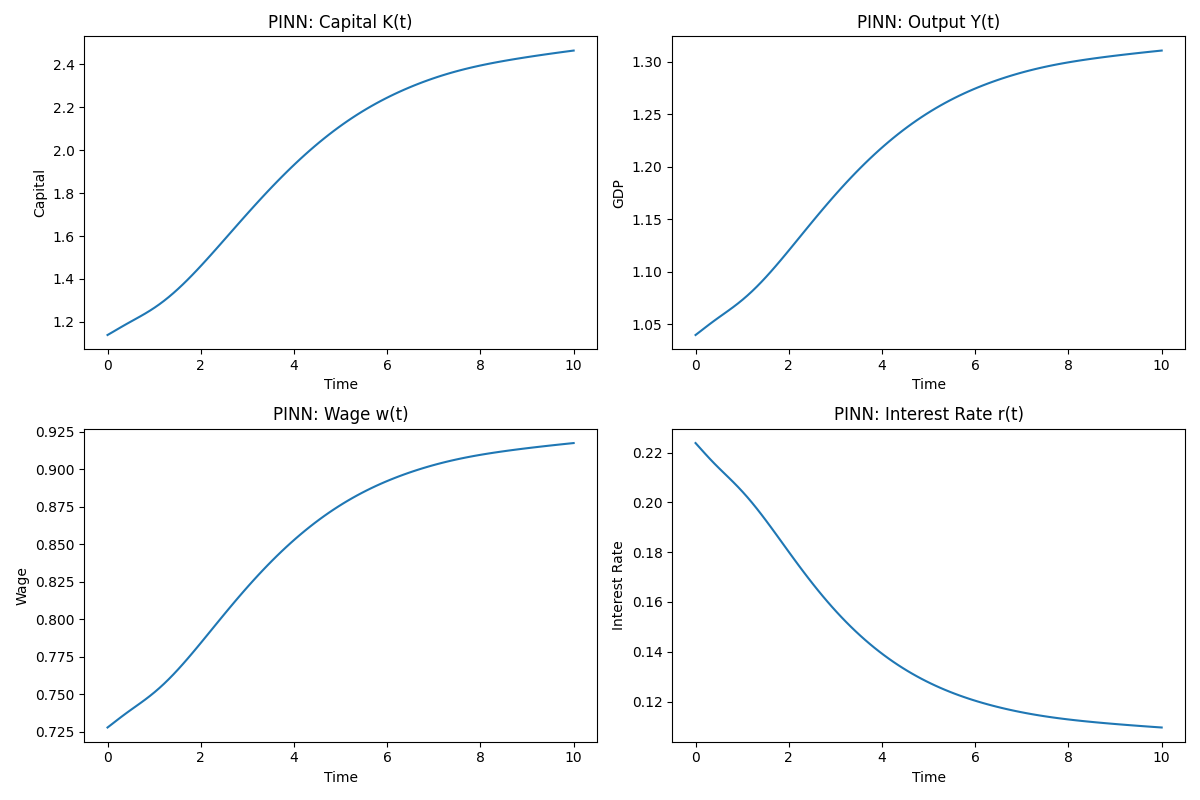}
    \caption{Computed time paths.}
    \label{fig:time-paths}
\end{figure}

The plots in Figure \ref{fig:time-paths} depict the evolution of capital, output, wages, and the interest rate over time. The top-left panel shows the evolution of capital, $K(t)$, indicating a steady increase in capital over time, starting from an initial value around $1.0$ and asymptotically approaching a higher steady-state level of $2.46$. This upward trend signifies capital accumulation. The top-right panel displays output, $Y(t)$, or Gross Domestic Product (GDP), mirroring the capital accumulation trend, output also rises over time. This positive relationship is consistent with standard economic theory. 

The bottom-left panel shows the trajectory of wages, $w(t)$. As the capital stock increases, the marginal product of labour also rises, leading to an increase in wages. The plot shows that wages grow steadily over the time period. This reflects the increasing productivity of labour due to a more capital-intensive economy. The bottom-right panel presents the behaviour of the interest rate, $r(t)$. In contrast to the other variables, the interest rate declines over time. As the capital stock grows, the marginal return to an additional unit of capital diminishes, causing the interest rate to fall to its steady state value of $0.1095$. This dynamic is a classic feature of the neoclassical growth model, where capital accumulation drives down the return on capital. 

\section{Conclusion and Discussion} \label{section:conclusion}

This paper proposes a mesh-free method, the ABH-PINN solver, for solving continuous-time heterogeneous agent macroeconomic models, using PINNs. By embedding the HJB and KF equations directly into the training objectives of neural networks, the ABH-PINN solver provides a potentially scalable and flexible alternative to traditional grid-based solvers. 

The use of PINNs opens promising avenues for addressing long-standing challenges in macroeconomic modelling. The method looks to address the curse of dimensionality, allowing future work to incorporate additional state variables such as multiple assets, aggregate risk, housing, or human capital without significant loss of tractability. Moreover, the smoothness and differentiability of the learned policy and distribution functions make PINNs particularly well-suited for applications in policy simulation, sensitivity analysis, and estimation.

The preliminary results have shown the ability of PINNs to produce economically plausible results. We plan to extend these results to the economic setup in \cite{achdou2022income} and compare our results to the FD solver, as well as test the robustness of this method and the computational burden as we increase additional state variables. An important direction for future work is to incorporate empirical data and evaluate the ability of the model to match observed wealth distributions and consumption responses. Incorporating government policy, aggregate shocks, portfolio choice, and more realistic liquidity constraints would enable the model to better match empirical data and provide deeper policy insights. In addition, integrating shape constraints via input-concave neural architectures could improve training stability and convergence.

Beyond the canonical ABH framework, this method provides a template for solving a wider class of economic models governed by complex PDE systems. As such, we are excited by the potential for the ABH-PINN solver to contribute both to computational economics and to the growing literature on scientific machine learning, demonstrating how structure-aware neural networks can advance economic modelling at the frontier of theory and computation.

\section*{Acknowledgments}
The author gratefully acknowledges the contributions of Tomas Kreuzinger, whose Master's thesis at the University of Cambridge laid the initial groundwork for this research. 

\bibliography{bib}

\end{document}